**Self-enhanced mobility enables vortex pattern formation in living matter**

Haoran Xu [1], Yilin Wu [1*]

[1] *Department of Physics and Shenzhen Research Institute, The Chinese University of Hong Kong, Shatin, NT, Hong Kong, P.R. China.*

*To whom correspondence should be addressed.  Mailing address: Room 306, Science Centre, Department of Physics, The Chinese University of Hong Kong, Shatin, NT, Hong Kong, P.R. China.  Tel: (852) 39436354.  Fax: (852) 26035204.
Email: ylwu@cuhk.edu.hk

**Abstract**:

Emergence of regular spatial patterns is a hallmark in living matter ranging from subcellular organelles to developing embryos and to ecosystems.  Mechanisms for the formation of ordered spatial patterns in biology often require chemical signaling that coordinates cellular behavior and differentiation.  Here we discovered a novel route to large-scale regular pattern formation in living matter mediated by purely physical interactions.  We found that dense bacterial living matter spontaneously developed an ordered lattice of mesoscale, fast-spinning vortices each consisting of ~$10^4$-$10^5$ motile cells; these mesoscale vortices were arranged in space over centimeter scale with apparent hexagonal order, while individual cells in the vortices moved in coordinated directions with strong polar and vortical order.  Single-cell tracking and numerical simulations suggest that the phenomenon is enabled by self-enhanced mobility of individual cells in the system.  Our findings demonstrate a simple physical mechanism for self-organized pattern formation in living systems and more generally, in other active matter systems near the boundary of fluidlike and solidlike behaviors.

**Introduction**

Ranging from subcellular organelle biogenesis to embryo development, the formation of self-organized structures is a hallmark of living systems. Emergence of ordered spatial patterns in biology can be driven by guidance of pre-patterned chemical fields (i.e., morphogens) [1,2], by chemotactic movement of cells [3,4], or by density-dependent motility coupled to cell-cell communications [5,6]. These mechanisms often require intricate chemical signaling that coordinates cellular behavior and even differentiation. Purely physical interactions can drive the formation of regular biological patterns such as crystalline vortex arrays in suspensions of sperms [7] and bacteria [8], but the characteristic length scale of such living patterns is on the order of individual cells. Here we discovered a novel route to self-organized pattern formation driven by physical interactions, which creates large-scale regular spatial structures with multiscale ordering. Specifically, we found that dense bacterial living matter spontaneously developed a lattice of mesoscale, fast-spinning vortices; these vortices each consisted of ~$10^4$-$10^5$ motile bacterial cells and were arranged in space over centimeter scale with apparent hexagonal order, while individual cells in the vortices moved in coordinated directions with strong polar and vortical order. Single-cell tracking and numerical simulations suggest that the phenomenon is enabled by self-enhanced mobility in the system, i.e., the speed of individual cells increasing with cell-generated collective stresses at a given cell density. Stress-induced mobility enhancement and fluidization is prevalent in dense living matter at various length scales, such as developing embryos [9,10], microbial biofilms [11], and cell cytoplasm packed with motor-driven macromolecules [12]. Our findings demonstrate that self-enhanced mobility offers a simple physical mechanism for pattern formation in living systems and more generally, in other active matter systems [13,14] near the boundary of fluidlike and solidlike behaviors [15-21].

**Results**

Our work began by investigating the patterns of active turbulence in dense active fluids [22]. Active turbulence is characterized by the formation of transient vortices that emerge and disappear randomly. It has been an intriguing question whether the active turbulence could spontaneously stabilize into ordered spatial structures, despite a number of theoretical predictions [21,23-25]. To explore such possibility, we chose to work with living active fluids consisting of motile bacteria. Swimming bacteria are premier systems for active matter studies [26]. We deposited dense suspensions (~8 × $10^{10}$ cells/mL) of the model flagellated bacterium *Serratia marcescens* (~2 µm in length and 0.8 µm in width) onto the surface of nutrient agar, and spread them into centimeter-scale, quasi-two-dimensional suspension films of a thickness ~5-10 µm (Methods). To our surprise, we discovered that the quasi-2D dense bacterial suspension films quickly developed an ordered spatial pattern over centimeter scale that was highly stable (Fig. 1a-c; Fig. S1; Video S1). The striking pattern consists of mesoscale vortices of a uniform size; e.g., the vortex diameter in Fig. 1a was 278±51 µm (mean±S.D., N=103; Methods). The vortices are exclusively clockwise (viewed from above the agar surface; N>30 experiments) and have an average collective speed (i.e., the magnitude of collective velocity measured by particle image velocimetry, corresponding to the ensemble average of the apparent single-cell speed measured in the dense suspension; Methods) of ~24 µm/s (Fig. 1b,c), which is ~2-fold higher than that outside the vortices (~12 µm/s). At the global scale, the pair-correlation function of the vortices shows a pronounced first-order peak corresponding to the average nearest vortex distance (Fig. 1d; Methods), while the triplet-distribution function that characterizes the hexagonal symmetry of a lattice [7,27] is peaked at $\frac{\pi}{3}, \frac{2\pi}{3}$ and $\pi$ (Fig. 1e; Methods). These results reveal hexagonal order of the vortex distribution, which is corroborated by the anisotropic six-fold symmetry in the 2D spatial correlations of the collective velocity field (Fig. 1f; Methods) and the vorticity field (Fig. S2; Methods).

Hexagonal vortex lattice patterns emerging from active turbulence in dense active fluids were previously predicted in theory [21,23] but never observed in experiment. To further characterize the phenomenon, we focused on the scale of individual vortices (Fig. 2a). Using fluorescent cells labeled by genetically encoded green fluorescent protein (GFP), we tracked the motion of single cells in the vortex pattern (Fig. 2b; Methods). Consistent with the results of collective velocity measurement, the mean speed of single cells in the vortices (24.9±13.4 µm/s; mean±s.d., N>5000) is higher than elsewhere (14.8±10.6 µm/s; N>5000) (Fig. 2c, upper). Interestingly, the motion of cells displays high local polar order inside vortices but is disordered outside the vortices (Fig. 2c, lower; Fig. 2d). We noted that cells could travel across the vortex boundaries (Fig. 2b). The cellular exchange across vortex boundaries demonstrates that there are no confinement walls at vortex boundaries, and thus the mechanism of vortex development is different from those arising in spatial confinement [28] or due to spatial guidance [29]. In addition, we used fluorescence intensity of cells as a measure of bacteria density and found that the cell density in the vortices was slightly lower than elsewhere by ~10-20% (Fig. S3; Methods), which accounted for the different brightness of vortices under phase contrast microscopy (Fig. 1a). Nonetheless, such density difference is not necessary for vortex formation, as some vortex lattice patterns had similar cell density across vortex boundaries (Fig. S3c-e).

We then followed the onset and development dynamics of the ordered vortex lattice (Fig. 3a; Video S2). We found that, after a short incubation time, an initially homogeneous bacterial suspension film displayed disordered collective motion reminiscent of bacterial turbulence [22], with transient high-speed streams and irregular vortices emerging at random locations and lasting a few seconds. The lifetime of the transient vortices gradually increased (Fig. 3b) and their shapes became more and more regular. Within ~5-10 min, the vortices became stabilized and settled into the hexagonal lattice pattern. Meanwhile, the collective speed of cells (Fig. 3c) shifted from unimodal to bimodal, with cells in vortices having higher collective speed. Notably, the emergence of ordered vortex lattice with a prolonged vortex lifetime and a bimodal speed distribution requires a critical cell density of ~5.5 × $10^{10}$ cells/mL in the bacterial suspension (Fig. 3d,e). Below this critical cell density, the bacterial suspension displayed active turbulence (Fig. S4), in which cells moved in locally coherent jets and domains, with transient vortices emerging occasionally but unable to stabilize.

Taken together, the dense bacterial active fluid spontaneously developed a large-scale vortex lattice pattern with multiscale ordering, including polar order at the single-cell level, mesoscale vortical order, and macroscopic hexagonal order. We note that the emergence of ordered vortex lattice does not involve cell-secreted polymers such as extracellular DNA and amyloid fibers (Methods), which have been shown to promote large scale self-organization in dense bacterial active matter [30,31]. The emergence of the ordered vortex lattice appears to require a sufficiently high motility of cells. Indeed, when we reduced the motility of *S. marcescens* by ~50% via violet light illumination (Fig. S5; Methods) [31], most vortices in the large scale vortex lattice dissolved, leaving few of them in the field of view (Video S3); meanwhile, the bimodal distribution of collective speed relaxed to unimodal (Fig. 3f). Moreover, formation of similar mesoscale vortices was observed with *Escherichia coli* (Fig. S6a,b; Video S4; Methods). However, the vortices that emerge in a suspension film of *E. coli* was sporadic (typically <10 in the field of ~10 $mm^2$) and disordered, presumably due to lower motility of *E. coli* cells; isolated *E. coli* cells have a mean speed of ~10 µm/s versus ~25 µm/s for isolated cells of *S. marcescens* (Fig. S5b).

What is then the origin of the ordered vortex lattice pattern? Before answering this question, we first sought to understand why individual cells in the vortices moved at a higher speed (Fig. 2c, upper), despite that all cells in the bacterial suspension should have a similar self-propulsion force because they were prepared from a homogeneous culture (Fig. S5b). This fact suggests that the physical environment of the vortices is different from elsewhere. We reasoned that the higher polar order of cellular motion in the vortices allows a coherent group of cells to generate a stronger local collective active stress (or active force density) [32], which would modify local mechanical environment and result in higher active transport for cells in the friction-dominated quasi-2D suspension film [30,33]. To examine this idea, we measured the relation between the apparent single-cell speed (denoted as $v$) and local polar order (denoted as $P = |\sum_i \boldsymbol{p}_i/n|$, where $n$ is the total number of cells near a cell of interest and $\boldsymbol{p}_i$ represents the polarization of cell $i$ among the $n$ cells); $P$ serves as a proxy of the local collective active stress $\sigma$ because $-\nabla \cdot \sigma = \sum_i \boldsymbol{\mathcal{F}}_i$ [32], where $\boldsymbol{\mathcal{F}}_i$ is the active force density generated by cell $i$ and $\boldsymbol{\mathcal{F}}_i \propto \boldsymbol{p}_i$. Indeed, we found that the apparent single-cell speed $v$ increases with local polar order and the relation can be fitted by a linear relation $v = \gamma_0(1 + \beta P)$, with $\beta$ being a positive dimensionless coefficient and $\gamma_0$ being the average speed of cells in the disordered phase with $P = 0$ (Fig. 4a). Interestingly, the coefficient $\beta$ increases with cell density and the value can be up to ~1.7 (Fig. 4b). Denoting the self-propulsion force of cells as $f_0$ and the single-cell mobility coefficient as $\mu$, we have $v = \mu f_0$ and the above linear relation implies the single-cell mobility coefficient $\mu \sim (1 + \beta P)$, which supports the notion that collective active stress in dense bacterial suspensions enhances the mobility of cells. We call this behavior as self-enhanced mobility, i.e., the speed of individual cells increasing with cell-generated collective stresses while holding cell density constant. Here mobility refers to the ability of cells to move under the mechanical constraints of the surroundings; by contrast, motility is the intrinsic self-propulsion speed of cells in an isolated environment, as measured in Fig. S5b.

Stress-induced enhancement of microscopic mobility (but not motility) is akin to stress-induced fluidization in passive colloidal glasses [34] and has been observed in other dense living matter systems [9-12]. The self-enhanced mobility of bacteria in dense suspensions was implied in the nonlinear positive relation between local polar order and *collective* speed in earlier studies on the phenomenon so-called "zooming bionematic phase" [35,36]; here our results based on single-cell speed measurement provide insight into this phenomenon. The behavior of self-enhanced mobility does not necessarily imply that the apparent speed of individual cells is higher than that of isolated cells at zero-density limit, because the average speed of cells in the disordered phase [i.e., $\gamma_0$ in the speed-order relation $v = \gamma_0(1 + \beta P)$] is inversely correlated with cell density. Also it should be distinguished from the macroscopic viscosity reduction observed in semi-dilute bacterial suspensions [37,38]; the latter arises from the onset of collective motion of cells under weak external shear, with the single cell speed remaining constant [39].

The self-enhanced mobility in dense bacterial suspensions offers a potential mechanism to explain the emergence of mesoscale vortices. Spontaneous fluctuations in the dense bacterial suspension could give rise to domains with higher polar order (hence with higher collective stress); such domains, with higher speed and longer persistence time than less-ordered domains, could recruit nearby cells via polar-alignment interactions and grow in size until cell recruitment is balanced by cell loss, eventually developing into stable vortices. To examine this hypothesis, we develop a Vicsek-type particle-based model [40] (Methods). Each bacterial cell $i$ is modeled as a Brownian particle that propels itself in 2D space with a variable mobility $\mu_i \equiv v_i/f_i$ ($v_i$ and $f_i$ are particle speed and self-propulsion force, respectively) that depends on local polar order (denoted as $P_i$) in the form of $\mu_i \propto (1 + \beta P_i)$ ($\beta \geq 0$); here $\beta$ corresponds to the dimensionless coefficient in the experimental relation between single cell speed and local polar order $v \sim (1 + \beta P)$, and hereinafter it is referred to as the mobility enhancement coefficient.

Numerical simulation of the particle-based model revealed that the emergent dynamics of the system depends on the mobility enhancement coefficient $\beta$ and particle activity (i.e., ensemble average of particle self-propulsion force $f_0 = \langle f_i \rangle$, proportional to the single-cell motility measured in experiments such as in Fig. S5b), as summarized in the phase map presented in Fig. 4c. The most notable feature of the phase map is that the system develops a stable vortex lattice pattern at sufficiently large $\beta$ and particle activity (Fig. 4d,e; Fig. S7, Fig. S8, Fig. S9; Video S5). The stable vortex pattern displays similar features to the large-scale vortex lattice observed in experiments (Fig. S7, Fig. S9), albeit with a weaker global hexagonal order (Fig. S8a). At low mobility enhancement coefficients (e.g., $\beta < 0.5$ for intermediate $f_0 = 10$), the model produces an active turbulence state without stable vortices, similar to the experimental phenomenon observed below the critical cell density for developing vortex lattices (as shown in Fig. S4). It is important to note that the mechanism of self-enhanced mobility is essential to produce the vortex lattice pattern in the model. Indeed, the system cannot develop stable vortices in the absence of mobility enhancement (i.e., $\beta = 0$), even when the activity is large; particles in the model tend to move in curved trajectories but it is insufficient to form stable vortices via polar alignment, although this tendency appears to control the size of vortices in the developed vortex pattern (Fig. S9k; Methods). These results support the notion that self-enhanced mobility underlies the onset of the vortex lattice pattern in dense bacterial suspensions.

Although the particle-based model can reproduce the formation of a vortex-lattice pattern, the hexagonal order of the vortex pattern is weaker compared to that observed in experiments. This is likely due to the absence of hydrodynamic interaction between vortices that could re-organize the spatial distribution of vortices. To account for the effect of hydrodynamic interaction during vortex lattice development, we adopt the framework of a minimal continuum model based on the seminal Toner-Tu active fluids model [41] and a Swift-Hohenberg-type fourth-order term [33] that successfully describes active turbulence in dense bacterial suspensions [22,23,42] (Methods). A key new ingredient in our model is that, in order to account for the behavior of self-enhanced mobility, the effective viscosity $\eta_{\text{eff}}$ (which is inversely proportional to single-cell mobility) entering the model parameters [43] is taken to depend on local polar order $P$ as $\eta_{\text{eff}} \propto 1/(1 + \beta P)$; here $\beta$ corresponds to the mobility enhancement coefficient described earlier in experiments and particle-based modeling (Methods). As shown in the phase map (Fig. 4f), numerical simulations of the continuum model produce a large-scale crystalline vortex lattice with hexagonal order and uniform chirality at strong mobility enhancement and high activity (denoted by the activity parameter $|S|$; $S < 0$ for pusher-type swimmers such as *S. marcescens* [22]); e.g., see Fig. 4g,h (also see Fig. S8, Fig. S9, Fig. S10 and Video S6). By contrast, the model produces active turbulence at weak mobility enhancement and low activity (e.g., Fig. S10a,b). These results are consistent with the experiments and particle-based simulations. We note that for sufficiently large activity $|S|$, the original model without self-enhanced mobility (i.e., equivalent to $\beta = 0$ in our simulation) was shown to produce hexagonal vortex lattices owing to nonlinear energy transfer [21]. However, our numerical simulations were performed in the regime of $S$ that would have led to the active turbulence state in the absence of self-enhanced mobility (Fig. 4f), and therefore, the vortex lattice development enabled by self-enhanced mobility as we studied here is via a different pathway, despite having similar phenomenology [21].

**Discussion**

Taken together, we have discovered that self-enhanced mobility in dense bacterial suspensions creates multiscale spatial order out of active turbulence, driving the formation of mesoscale vortices arranged in a centimeter-scale hexagonal lattice. Vortex structures are characteristic of turbulent flows that are ubiquitous in nature spanning vast length scales, ranging from quantum fluids [44] to galaxies [45]. Here we demonstrate a unique route via which chaotic vortex structures in turbulent flows can spontaneously stabilize into large-scale ordered vortex lattices. The stabilization of transient vortices in our study does not require spatial confinement [46] or geometrical guidance [29]. Ordered vortex structures, such as hexagonal vortex lattice and square vortex lattice states, have been theoretically predicted in dense active polar fluids for a long time [21,23,24,33,47,48]. Our work provides the first experimental observation of such ordered vortex structures in dense bacterial suspensions emerging from active turbulence via the stabilization of chaotic vortices. Crystalline vortex arrays were also reported in suspensions of sperms [7] and bacteria [8] via self-assembly processes due to hydrodynamic entrainment between nearby cells, and the characteristic length scale of these living patterns is on the order of individual cells. By contrast, the relative length scale of the vortices uncovered here (several hundred micron) compared to single-cell size (a few micron) is ~two orders of magnitude greater. Lattice-like vortex patterns were also reported in active filaments [49] and in filamentous living organisms [50,51]. Nonetheless, the phenomena in these examples do not display apparent hexagonal order, and they are due to steric interactions driving spatial segregation and clustering of self-propelled units, which is fundamentally different from our finding both at the mesoscale and the microscale.

More generally, our work suggests self-enhanced mobility as a generic mechanism for self-organization in dense living matter and synthetic active matter [52,53]. In this mechanism, the mobility enhancement could be due to modification of local physical environment by self-generated mechanical stresses (via motility or growth), or due to active regulation of motility in response to local stimuli. The mechanism may therefore provide new insights into the formation of stable vortex structures in cytoplasmic flows [54] as well as in large-scale cell movement of developing animal embryos, such as whirling somitomeres alongside the neural tube [55,56] and vortex-like polonaise movements during primitive streak formation [57]. It may also inform the design of spatial patterns in synthetic active matter for desired functionalities. Meanwhile, active fluids share phenomenological similarities with quantum superfluids, such as the formation of low-viscosity modes and coherent vortex structures [58]. Our finding of vortex pattern formation enabled by mobility enhancement may offer a new perspective to examine the analogy between active and quantum fluids.

**Methods**

**Bacterial culture, microscopy and image processing.** The following strains were used: wildtype *Serratia marcescens* ATCC 274; *S. marcescens* GFP (*S. marcescens* ATCC 274 with constitutive expression of green fluorescent protein encoded on the plasmid pAM06-tet [59] from Arnab Mukherjee and Charles M. Schroeder, University of Illinois at Urbana-Champaign); *E. coli* HCB1737 (a derivative of *E. coli* AW405 with wildtype flagellar motility; from Howard Berg, Harvard University, Cambridge, MA); *E. coli* GFP (*E. coli* HCB1737 hosing the plasmid pAM06-tet). Plasmids were transformed into the strains by electroporation. Single-colony isolates were grown overnight (~13–14 h) with shaking in LB medium (1% Bacto tryptone, 0.5% yeast extract, 0.5% NaCl) at 30 °C to stationary phase. For *S. marcescens* GFP and *E. coli* GFP, kanamycin (50 µg/mL) was added to the growth medium for maintaining the plasmid. All imaging was performed on a motorized inverted microscope (Nikon TI-E). The phase-contrast images were acquired with a 4×, 10× or 20× objective; the images were recorded by a scientific complementary metal-oxide-semiconductor (sCMOS) camera (Andor Zyla 4.2 PLUS USB 3.0) at 30 or 10 fps (for recording the onset and development of the vortex lattices) and at full frame size (2048×2048 pixels).

Microscopy images were processed using the open-source Fiji (ImageJ) software (http://fiji.sc/Fiji) and custom-written programs in MATLAB (The MathWorks; Natick, Massachusetts, United States). To compute the collective velocity field in the quasi-2D bacterial suspension films, we performed particle image velocimetry (PIV) analysis based on phase-contrast time-lapse videos using an open-source package MatPIV 1.6.1 written by J. Kristian Sveen (http:// folk.uio.no/jks/ matpiv/index2.html). The PIV analysis yielded space- and time-dependent collective velocity field $\vec{v}(\vec{r},t) = (v_x, v_y)$ and vorticity field $\omega(\vec{r},t) = \partial_x v_y - \partial_y v_x$. The magnitude of $\vec{v}(\vec{r},t)$ is taken as the collective speed. Positive and negative values of $\omega(\vec{r},t)$ correspond to counterclockwise and clockwise rotation, respectively. To visualize the collective velocity or vorticity field computed by PIV analysis, the velocity or vorticity field was coarse-grained and plotted on a square mesh with appropriate grid spacing. Flow streamlines were computed by the built-in streamslice function in MATLAB and plotted on the vorticity field. The trajectories of bacterial cells were obtained by single-cell tracking based on the recorded fluorescence videos, using a custom-written program in MATLAB [60]. The steady-state local polar order of tracked single cells located near position $\vec{r}$ as shown in Fig. 2d was defined as $P(\vec{r}) = |\langle \vec{n}_i \rangle_{i,t}|$, where $\vec{n}_i = \vec{v}_i/|\vec{v}_i|$ is the velocity direction and the angular bracket represents both averaging over the cell index $i$ whose position $\vec{r}_i$ satisfies $|\vec{r} - \vec{r}_i| \leq 6.5\ \mu m$ and averaging over the entire time window of cell tracking.

**Particle-based simulation and data analysis.**
In the simulations, we considered a collective of overdamped self-propelled particles moving in 2D space. Specifically, for the $i$-th particle, its self-propulsion speed is given by $v_i = \mu_i f_i$, where $\mu_i$ is the effective mobility coefficient of the particle and $f_i$ is the self-propulsion force. The mobility $\mu_i$ follows a linear relation with the local polar order in the vicinity of particle $i$ (denoted as $P_i$) in the form of $\mu_i \propto (1 + \beta P_i)$; here $\beta \geq 0$ is the dimensionless mobility enhancement coefficient. In addition, the model accommodates steric and hydrodynamic interactions between swimmers as effective coupling of particle orientation and angular velocity, in a way similar to that adopted earlier to simulate the collective motion of millions of cells in large-scale dense bacterial active fluids [61]. The dynamics of the $i$-th particle in terms of its center of mass position $\vec{r}_i$, orientation $\theta_i$, and angular velocity $\omega_i$ is governed by the following equations:

$$\dot{\vec{r}}_i = \mu_i f_i \hat{n}_i + \sqrt{2D_r}\ \xi_r, \qquad [1]$$

$$\dot{\theta}_i = \frac{k_\theta}{m_i} \sum_j sin(\theta_j - \theta_i) + \omega_i + \sqrt{2D_\theta}\, \xi_\theta, \qquad [2]$$

$$\dot{\omega}_i = -\frac{\omega_i}{\tau} + \frac{k_\omega}{m_i} \sum_j (\omega_j - \omega_i) + \xi_b + \sqrt{2D_\omega}\, \xi_\omega. \qquad [3]$$

In Eq. [1], $\hat{n}_i$ is a unit vector of orientation $\theta_i$. The mobility $\mu_i$ of particle $i$ depends on the local polar order $P_i$ near the particle in the linear form of $\mu_i = \Gamma_0(1 + \beta P_i)$; $P_i$ ($\in [0,1]$) was calculated as $P_i = \langle |\sum_j \frac{\vec{r}_j}{|\vec{r}_j|}|/m_i \rangle_{t_0}$, where the summation was over $m_i$ particles found in the neighborhood of particle $i$ (with $|\vec{r}_j - \vec{r}_i| \leq 5$; including particle $i$ itself) and the angular bracket denotes averaging over a time $t_0 = 10$ to smooth out fluctuations. In Eq. [2,3], the summation was also over $m_i$ particles found in the neighborhood of particle $i$ (with $|\vec{r}_j - \vec{r}_i| \leq 1$). For the rest of the equations, $k_\theta$ represents the strength of effective polar alignment between velocity directions due to steric and hydrodynamic interactions between cells; $k_\omega$ represents the strength of diffusive coupling between angular velocities due to local fluid vorticity; $D_r$, $D_\theta$ and $D_\omega$ are translational, rotational and angular velocity diffusivity, respectively; $\xi_r$, $\xi_\theta$ and $\xi_\omega$ are Gaussian white noise with zero mean and unit variance; $\tau$ arises from the relaxation of local angular velocities; and a bias noise $\xi_b = sign(\omega_i)\exp(-|\omega_i|/\omega_0 \xi)$ where $sign(\omega_i)$ is the current sign of $\omega_i$, $\omega_0$ is a constant, and $\xi$ is a uniform noise in $[0,\eta_{bias}]$. The signed bias noise term together with the relaxation term in angular velocity dynamics in Eq. [3] allows for spontaneous chiral symmetry breaking [61]. Adding a finite negative bias $b$ to the particles' angular velocity (by substituting $\omega_i$ with $(\omega_i + b)$ in Eq. [2,3]; accounting for the swimming bias near a solid substrate [62] seen in Extended Data Fig. 2b) produced vortex lattice patterns with exclusively CW chirality as seen in experiments (see Supplementary Information).

For the $i$-th particle, its position $\vec{r}_i$, orientation $\theta_i$ and angular velocity $\omega_i$ evolved according to Eq. [1-3]. In Eq. [2,3], the coupling range for orientation and angular velocity dynamics was chosen as 1. For each particle, the self-propulsion force $f_i$ was sampled from a Gaussian distribution with the mean of $f_0$ and standard deviation of $0.4 f_0$ to account for the heterogeneity of cell motility. We chose the parameters in Eq. [3] that yielded the temporally averaged angular velocity $\Omega \equiv \langle \omega_i \rangle_{i,t} = -1$. At zero-density limit (i.e., without inter-particle interactions) the particles move circularly in either clockwise or counterclockwise manner, each following a Gaussian-like local curvature distribution of the same shape (i.e., standard deviation) and of a mean set by $|\Omega| \equiv \langle |\omega_i| \rangle_{i,t}$; note that the local curvature of the $i$-th particle's trajectory is proportional to the reorientation rate $\Delta \theta_i / \Delta t$ over a finite time based on Eq. [2]. All simulations were initialized with uniform random distributions of particle position, orientation and angular velocity (in the range of [-1, 1]). To convert numbers in particle-based simulations to physical units, we set one unit length in the model as 2 μm (that is, about one cell length), and one time unit (equivalent to 100 time steps in the simulation) as 1 s. We used particle number $N = 1600000$ and size of the system $L_x = L_y = 400$ (with periodic boundary conditions), so the particle density $\rho_0 = \frac{N}{L_x L_y} = 10$ corresponded to 2.5 cells/μm² in experiments. To obtain the phase map, the mean self-propulsion force $f_0$ were varied from 1 to 20, and the mobility enhancement coefficient $\beta$ were varied from 0.0 to 3.0 to be consistent with experimental results. For other simulation parameters, the following values were used: $\Gamma_0 = 1$, $k_\theta = 0.2$, $k_\omega = 2$, $D_r = 0.02$, $D_\theta = 0.02$, $D_\omega = 0.02$, $\tau = 10$, $\omega_0 = 0.6$, $\eta_{bias} = 1$. Both $\Gamma_0$ and $\beta$ in the expression of particle speed shall depend on particle density in accordance with the experimental measurement of cell speed. As we focus on the phenomenon of single-cell speed increasing with local polar order at a constant cell density in dense suspensions, the model has been designed to describe the emergent dynamics at a relatively high particle density, and the density dependence of model parameters is not considered here.

The collective velocity field $\vec{v}(\vec{r}, t)$ in particle-based simulations was calculated based on single-particle velocity $\vec{v}_i(\vec{r}, t)$ as $\vec{v}(\vec{r}, t) = (v_x(\vec{r}, t), v_y(\vec{r}, t)) = |\sum_i \vec{v}_i(\vec{r}, t)|/m$, where $m$ is the total number of particles whose position $\vec{r}_i$ satisfies $|\vec{r}_i - \vec{r}| \leq 2$. Similarly, the polar order field was defined as $P(\vec{r}, t) = |\sum_i \frac{\vec{v}_i(\vec{r}, t)}{|\vec{v}_i(\vec{r}, t)|}|/m$. The vorticity field $\omega(\vec{r}, t)$ was calculated based on the computed collective velocity field $\vec{v}(\vec{r}, t)$ as $\omega(\vec{r}, t) = \partial_x v_y(\vec{r}, t) - \partial_y v_x(\vec{r}, t)$. The particle density field $\rho(\vec{r}, t)$ was calculated as $\rho(\vec{r}, t) = \frac{N(\vec{r})}{\pi d^2}$, where $N(\vec{r})$ represented the number of particles whose position $\vec{r}_j$ satisfies $|\vec{r}_j - \vec{r}| \leq d$ ($d = 2$).

To obtain the phase map as shown in Fig. 4c, we distinguished the mode of emergent collective motion in particle-based simulations by the autocorrelation time of the vorticity field $\omega(\vec{r}, t)$. The autocorrelation function of vorticity field is defined as $C_t(\Delta t) = \frac{\langle \omega(\vec{r}, t) \omega(\vec{r}, t + \Delta t) \rangle_{\vec{r}, t}}{\langle \omega(\vec{r}, t)^2 \rangle_{\vec{r}, t}}$, where the angular brackets represent both averaging over the entire simulation domain and averaging over the time $t$ from 2000 to 3000 time units (also keeping $t + \Delta t$ in the same range). The autocorrelation function $C_t(\Delta t)$ was fitted to $C_t(\Delta t) \sim e^{-\Delta t/\tau_0}$, yielding the autocorrelation time $\tau_0$ of the vorticity field. If $\tau_0 \geq 10$ (corresponding to 10 s in experiments), the emergent motion mode was classified as the vortex lattice state; otherwise, the emergent motion mode was classified as active turbulence state.

We note that, inspired by the nonlinear positive relation between local polar order and collective speed in dense bacterial suspensions reported previously [36], a power-law type positive relation between mobility and local polar order $\mu_i \propto P_i^\beta$ ($\beta \geq 0$) could also produce qualitatively similar results in all particle-based simulations.

**Numerical simulation of continuum model.**
In our continuum model, the collective velocity field $\vec{v}$ of the quasi-2D bacterial active fluid is governed by the following equation:

$$\partial_t \vec{v} + \lambda_0 \vec{v} \cdot \nabla \vec{v} = (-\nabla G + \lambda_1 \nabla |\vec{v}|^2) - (\alpha + \varphi |\vec{v}|^2)\vec{v} + \Gamma_1 \nabla^2 \vec{v} - \Gamma_2 (\nabla^2)^2 \vec{v} + \kappa \omega \vec{v}. \qquad [4]$$

The core of the equation is the seminal Toner-Tu active fluids model [41] and a Swift-Hohenberg-type fourth-order term [33,63]. Here $G$ is pressure, and ($\alpha, \varphi, \lambda_1, \kappa, \Gamma_1, \Gamma_2, \lambda_0$) are model parameters with the latter three related to the effective viscosity of the active fluid $\eta_{\text{eff}}$ (which is inversely proportional to single-cell mobility $\mu$) [43]: $\Gamma_1 = -\gamma_1 \Gamma_0/\eta_{\text{eff}}$, $\Gamma_2 = \gamma_2 \Gamma_0/\eta_{\text{eff}}$, and $\lambda_0 = (1 - S\eta_0/\eta_{\text{eff}})$, with $\gamma_1, \gamma_2, \Gamma_0, \eta_0, \epsilon$ being constants; in the last expression $S$ is the activity parameter ($S < 0$ for pusher-type swimmers such as *S. marcescens*) [22] and $\eta_0$ is an effective zero-shear viscosity. The system is treated as an incompressible fluid (i.e., $\nabla \cdot \vec{v} = 0$). To account for the behavior of self-enhanced mobility, $\eta_{\text{eff}}$ is taken to depend on local polar order $P$ as $\eta_{\text{eff}} = \eta_0/(1 + \beta P)$ and it can be expressed in terms of $|\vec{v}|$ (noting that the collective speed $|\vec{v}| \propto \mu P = P/\eta_{\text{eff}}$). In addition, the vorticity-dependent term $\kappa \omega \vec{v}$ ($\omega$ is vorticity of the velocity field) allows chirality selection depending on the sign of parameter $\kappa$, which accounts for the chirality bias observed in experiments; note that the system displays spontaneous chiral symmetry breaking at $\kappa = 0$, i.e., the emergent vortex lattice pattern has equal probabilities of being CW or CCW.

The velocity field $\vec{v}(\vec{r}, t) = (v_x(\vec{r}, t), v_y(\vec{r}, t))$ in Eq. [4] was solved numerically using the pseudo-spectral method [64]. The vorticity field $\omega$ was calculated based on the computed velocity field $\vec{v}$ as $\omega = \partial_x v_y - \partial_y v_x$. The incompressibility condition $\nabla \cdot \vec{v} = 0$ was enforced by correcting the gradient of hydrodynamic pressure every time step using the method of Lagrange multiplier [65]. All equations were solved or integrated in Fourier space except for the computation of the nonlinear terms [65]. The velocity field was inversely transformed back to real space for visualization. The velocity field $\vec{v}$ evolved in 2D cubic grids with sizes $N_x \times N_y = 256 \times 256$ lattice points. Size of the system $L_x = L_y = 44\pi$ (with periodic boundary conditions) and time step $dt = 0.00001$ were used in the simulations. Mobility enhancement coefficient $\beta$ and activity $|S|$ were varied from 0.0 to 3.0 (to be consistent with experimental results) and from 0 to 3, respectively. All simulations were initialized with uniform random distributions of velocity direction in $[0, 2\pi]$ and velocity magnitude in $[0, 1]$. For other simulation parameters, the following values were used: $\alpha = 0.2$, $\varphi = 0.01$, $\gamma_1 = 2$, $\gamma_2 = 1$, $\Gamma_0 = 1$, $\eta_0 = 1$, $\epsilon = 1$, and $\kappa = -0.1$. The simulations of continuum model were written in C++ and performed on NVIDIA GeForce RTX 2080Ti through the application programming interface (API) of Compute Unified Device Architecture (CUDA; Nvidia Corporation; Santa Clara, California, United States) toolkit and platform.

Finally, similar to the particle-based simulations, we note that a power-law type relation between effective viscosity and local polar order $\eta_{\text{eff}} = \eta_0 P^{-\beta} \propto |\vec{v}|^{-\beta/(1+\beta)}$ ($\beta \geq 0$) could also produce qualitatively similar results in all continuum simulations.

**Supplementary Materials**, including Supplementary Videos, is available in the online version of the paper.

**Data availability.** The data supporting the findings of this study are included within the paper and its Supplementary Materials.

**Code availability.** The custom codes used in this study are available from the corresponding author upon request.

**Acknowledgements**. We thank Arnab Mukherjee and Charles M. Schroeder (University of Illinois at Urbana-Champaign) and Howard Berg (Harvard University) for generous gifts of bacterial strains. We thank Yi Wang (The Chinese University of Hong Kong) for help with computing resources, and Rui Zhang (HKUST) for helpful discussions on numerical simulations. This work was supported by the Ministry of Science and Technology of China (no. 2020YFA0910700), the Research Grants Council of Hong Kong SAR (RGC ref. nos. 14307821,14307822, RFS2021-4S04 and CUHK Direct Grants) and the National Natural Science Foundation of China (NSFC no. 31971182). Y.W. acknowledges support from New Cornerstone Science Foundation through the Xplorer Prize.

**Author Contributions**: H.X. discovered the phenomena, designed the study, performed experiments, developed the model, performed simulations, analyzed and interpreted the data. Y.W. conceived the project, designed the study, analyzed and interpreted the data. Y.W. and H.X. wrote the paper.

**Figures**

Figure 1

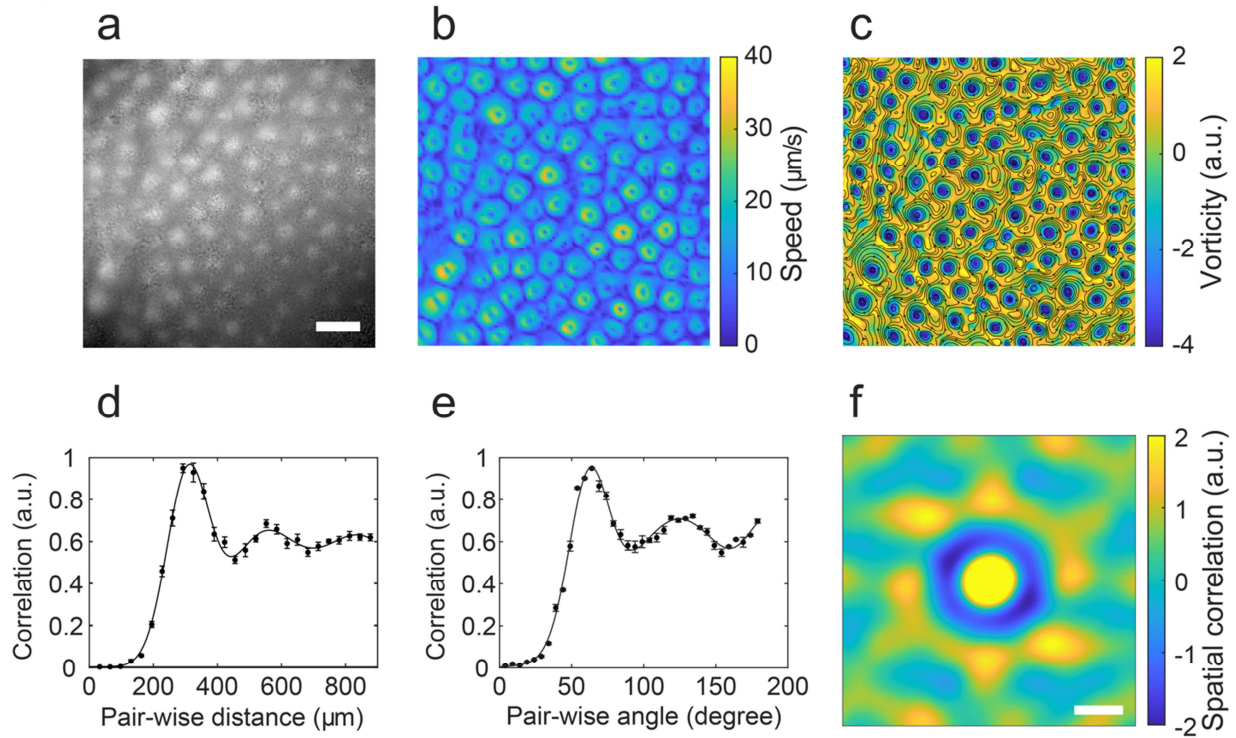

**Fig. 1. Ordered vortex lattice in quasi-2D dense bacterial active fluids.** (**a**) Phase-contrast image of the large-scale vortex lattice in a dense bacterial suspension consisting of *S. marcescens* (~8 × $10^{10}$ cells/mL). (**b**) Spatial distribution of time-averaged collective speed for panel a (Methods). Colorbar to the right is in unit of µm/s. (**c**) Spatial distribution of time-averaged vorticity for the pattern in panel a (Methods). The value of vorticity indicated by the colorbar is normalized by the mean of absolute vorticity in the entire field. Positive and negative values of vorticity correspond to counterclockwise and clockwise rotation, respectively. Lines ending with an arrow and overlaid to the vorticity colormap indicate the flow streamlines of the collective velocity field (Methods). Panels a-c share the same scale bar (500 µm). Data presented in panels b,c were averaged over a duration of 10 s. The vortex lattice exhibits exclusively clockwise chirality (over 30 experiments) viewed from above the agar plate (i.e., viewed from air to the liquid film). (**d,e**) Pair-correlation function (panel d) and triplet-distribution function (panel e) of the ordered vortex lattice pattern (Methods). Solid lines were obtained by fitting the data into a sum of three Gaussian functions (Methods). Error bars indicate S.E.M. (N=4). (**f**) 2D spatial correlation of the collective velocity field associated with panel a (Methods). Colorbar provided to the right is in arbitrary unit. Scale bar, 200 µm.

Figure 2

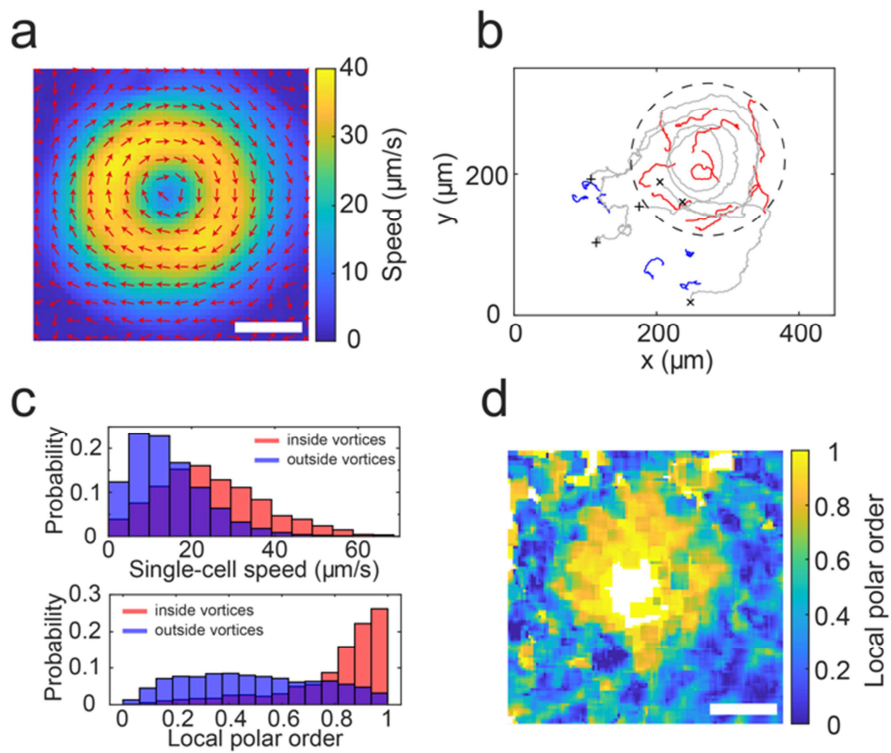

**Fig. 2. Microscale characteristics of the ordered vortex lattice**. (**a**) Time-averaged collective velocity field of a single vortex in the lattice (Methods). The data were averaged over a duration of 10 s. Arrows and colormap represent collective velocity directions and magnitude, respectively, with the colorbar provided to the right (in unit of μm/s). Scale bar, 100 μm. (**b**) Representative single-cell trajectories near a vortex in the lattice. Red and blue lines show the trajectories (lasting for 5 s) inside and outside the vortex, respectively, with the vortex boundary indicated by the dashed line; gray lines represent the trajectories (lasting for 25 s) of cells travelling across the vortex boundary (+, starting points; ×, ending points). (**c**) Probability distributions of single-cell speed (upper) and local polar order of single-cell motion (lower) in the vortex lattice. Red and blue histograms represent data obtained inside and out of the vortices, respectively. (**d**) Spatial distribution of local polar order near the vortex associated with panel b (Methods). Colorbar to the right is in arbitrary unit. Scale bar, 100 μm.

Figure 3

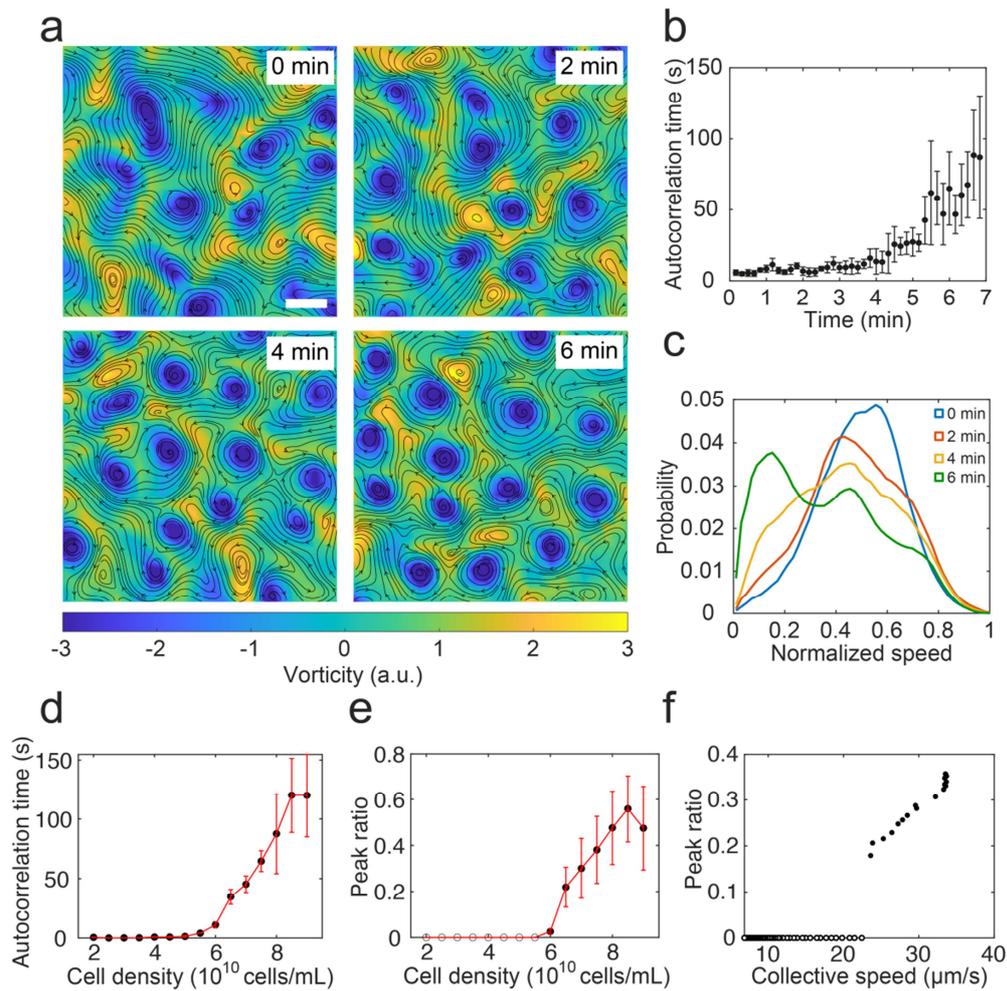

**Fig. 3. Emergence of ordered vortex lattice**. (**a**) Time sequence of instantaneous vorticity field during the emergence of an ordered vortex lattice in a dense bacterial suspension (~8 × $10^{10}$ cells/mL). The vorticity field is plotted in the same manner as in Fig. 1c. Scale bar, 200 µm. (**b**) Transient autocorrelation time of collective velocity (as a measure of vortex lifetime) during the developmental process of vortex lattice shown in panel a (Methods). Error bars indicate S.E.M. (N>=9; Methods). (**c**) Probability distributions of normalized collective speed during developmental process shown in panel a. Colors represent the elapsed time (blue: 0 min, red: 2 min, yellow: 4 min; green: 6 min). (**d,e**) Steady-state autocorrelation time of collective velocity (panel d) and peak ratio in collective speed distribution (panel e) plotted against cell density. The peak ratio is defined as the ratio between the heights of the higher-speed peak (corresponding to the vortex region) and the lower-speed peak in the collective speed distribution such as shown in panel c (Methods). The solid circles in panel e represent data from bimodal speed distributions, while the empty circles correspond to zero peak ratio and represent data from unimodal speed distributions (i.e., the higher-speed peak has a height of zero). (**f**) Peak ratio in collective speed distribution plotted against the average collective speed of cells (cell density: ~8 × $10^{10}$ cells/mL). Here the average collective speed serves as a proxy of single-cell motility (see Fig. S5), which was tuned by violet-light illumination (Methods).

Figure 4

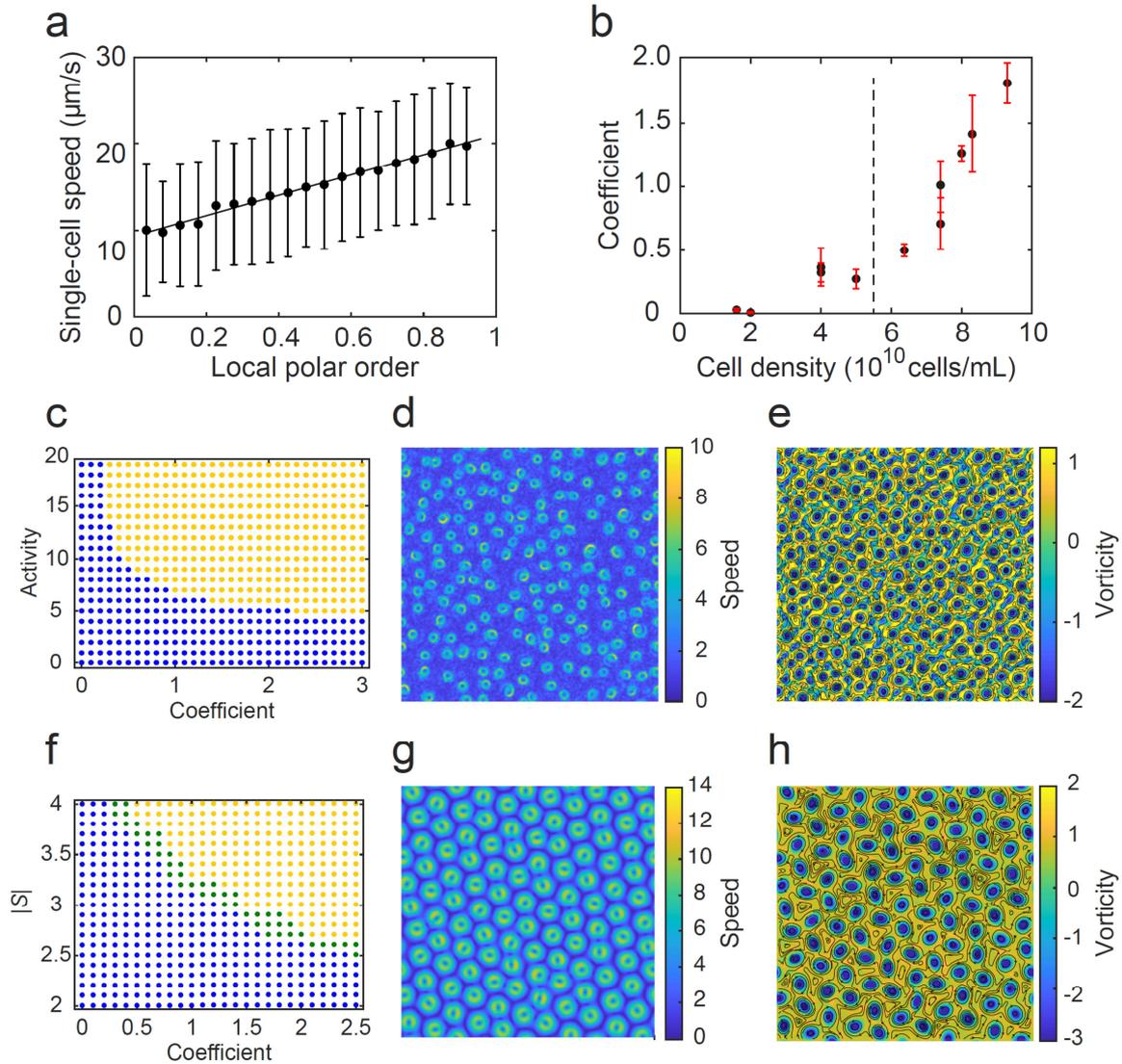

**Fig. 4. Self-enhanced mobility in dense bacterial suspensions underlies the formation of ordered vortex lattices.** (**a**) Apparent single-cell speed obtained by single-cell tracking plotted against local polar order in dense bacterial active fluids displaying ordered vortex lattices (cell density, ~8× $10^{10}$ cells/mL; Methods). Solid line is linear fit to the data in the form of $v \sim (1 + \beta P)$, with the dimensionless coefficient $\beta$ being 1.25±0.12; error bars represent standard deviation (N>500 single-cell trajectory segments for each data point). (**b**) Coefficient $\beta$ in the experimental relation between apparent single-cell speed and local polar order $v \sim (1 + \beta P)$ plotted against cell density. Dashed line indicates the minimal cell density for the development of ordered vortex lattices. Error bars represent fitting error of $\beta$. (**c**) Phase diagram of emergent collective motion patterns in particle-based simulations plotted in the plane of mobility enhancement coefficient $\beta$ and particle activity $f_0$ (yellow dots: vortex lattice; blue dots: active turbulence). (**d,e**) Spatial distributions of time-averaged collective speed (panel d) and vorticity (panel e) in the vortex lattice pattern produced in a representative particle-based simulation at steady state (Methods). Data presented in the two panels were averaged over a duration of 100 time units and plotted in the same manner as Fig. 1b,c. Simulation parameters: $f_0 = 10$ and

$\beta = 1.2$. (**f**) Phase diagram of emergent collective motion patterns in continuum modeling plotted in the plane of mobility enhancement coefficient $\beta$ and activity $|S|$ (yellow: hexagonal vortex lattice; blue: active turbulence; green: coexistence of hexagonal vortex lattice and active turbulence). (**g,h**) Spatial distributions of time-averaged collective speed $|\vec{v}|$ (panel g) and vorticity $\omega$ (panel h) in the vortex lattice pattern produced in a representative continuum simulation at steady state. Data presented in the two panels were averaged over a duration of 10 time units and plotted in the same manner as Fig. 1b,c. Simulation parameters: $|S| = 3.5$ and $\beta = 2.0$.

**Supplementary Figures**

**Fig. S1**

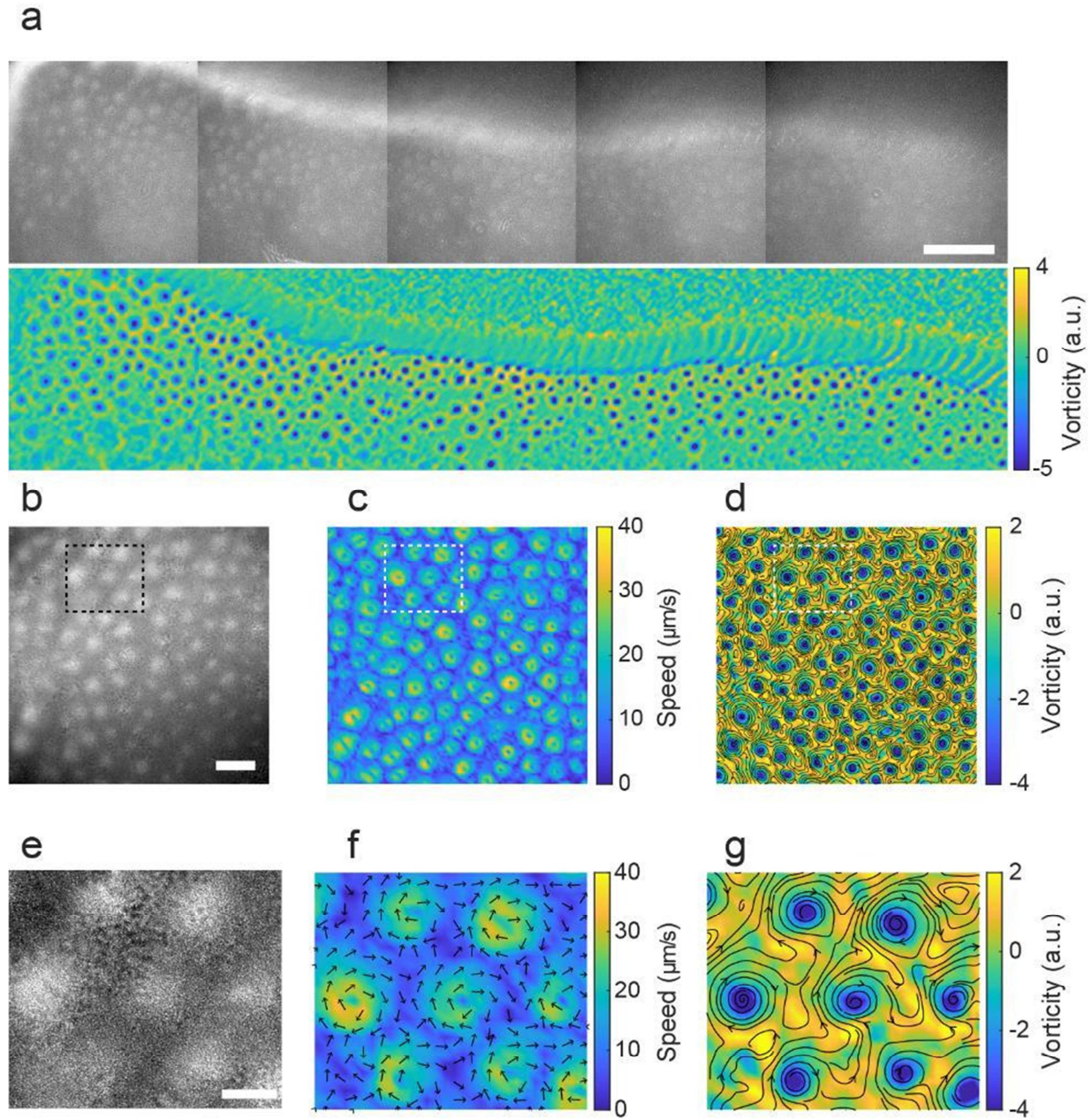

**Fig. S1. Vortex lattice pattern at different length scales**. (**a**) Vortex lattice pattern at centimeter scale. Upper: phase-contrast image; lower: vorticity field associated with the phase-contrast image, computed based on the collective velocity field. The length and width of the panel is ~1.4 cm and 3.3 mm, respectively. The upper panel is a composite image obtained by stitching a sequence of 5 images taken at connected smaller windows; for each window, a short video lasting ~10 s was taken to compute the collective velocity field via PIV (Methods). The vortices appeared everywhere in the image except at the upper region; this region is close to the edge of the suspension film, where the fluid film has a greater thickness. The vortices at the rightmost region in the phase-contrast image (upper panel) are not apparent because the cell densities inside and outside the vortices are similar, but they can be visualized in the vorticity field (lower panel). Scale bar, 1 mm. (**b-d**) Phase-contrast image (panel c), collective speed distribution (panel d), and vorticity distribution (panel e) of the vortex lattice pattern shown in main text Fig. 1a-c. Panels b-d share the same scale bar (500 µm). (**e-g**) Enlarged view of the area enclosed by the box in panels b-d. The arrows in panel f represent the velocity direction of the local collective velocity. Scale bar, 200 µm.

**Fig. S2**

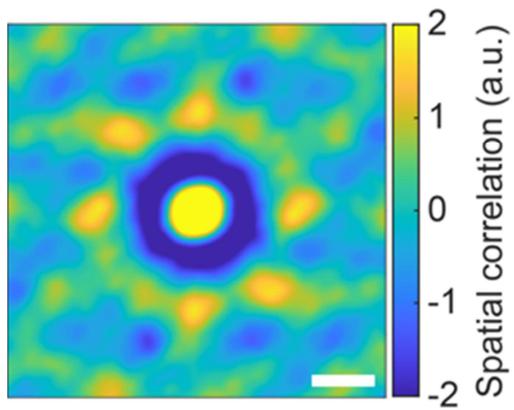

**Fig. S2. Spatial correlation of the vorticity field in a quasi-2D dense bacterial active fluid.** This figure is associated with main text Fig. 1a-c (Methods). Colorbar provided to the right is in arbitrary unit. Scale bar, 200 µm.

**Fig. S3**

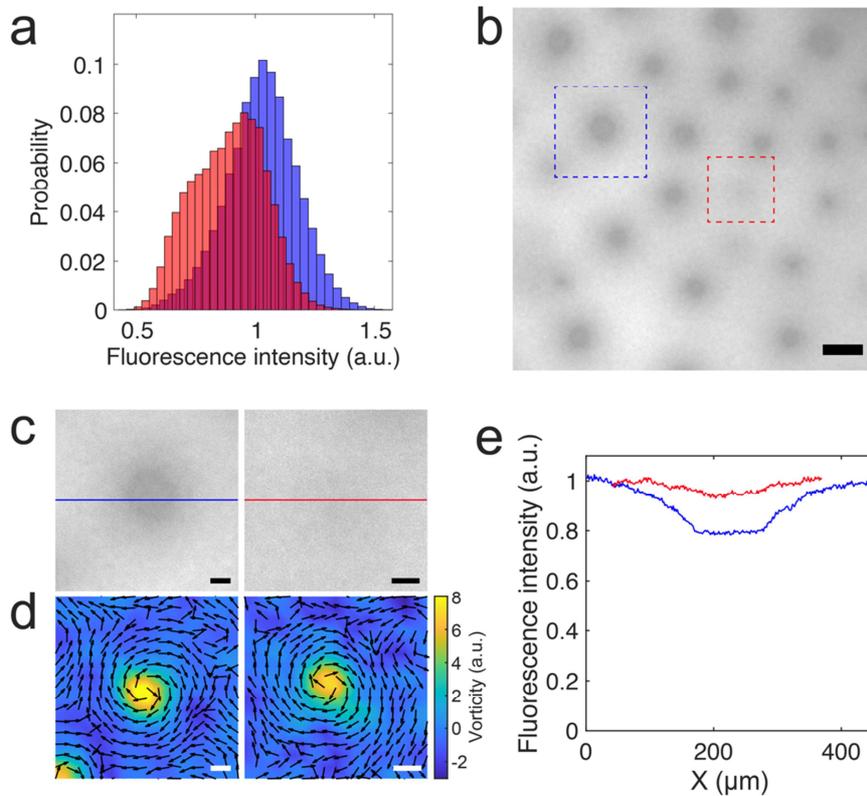

**Fig. S3. Characterization of cell density distribution in the vortex lattice pattern**. (**a**) Probability distributions of fluorescence intensity inside (red) and outside (blue) the vortices in a representative vortex lattice pattern. Fluorescence intensity is a measure of cell density because cells were labelled by GFP. The averaged density difference between inside and outside vortices of the vortex lattice pattern obtained from the fluorescence intensity distributions is ~ 15%. (**b**) Spatial distribution of the fluorescence intensity in the vortex lattice pattern analyzed in panel a. The cell density inside vortices is slighter lower than or comparable to the regions outside the vortices. Scale bar, 200 μm. (**c**) Enlarged view of the areas enclosed by the boxes in panel b. Each box contains a vortex. Left (blue box in panel b): the cell density inside the vortex is slightly lower than outside; right (red box in panel b): the cell density inside the vortex is similar to outside. (**d**) Vorticity fields corresponding to the regions shown in panel c. Color bar represents the magnitude of vorticity while the arrows represent the direction of local collective velocity vectors. The vorticity fields for the two vortices are similar, suggesting that cell density difference is not necessary for vortex formation. Scale bars in panel c and d, 50 μm. (**e**) Fluorescence intensity plotted along the straight lines across the two vortices in panel c. The red and blue plots in panel e were plotted along the lines with the respective color in panel c. Data shown in the plots were normalized by the mean fluorescence intensity in regions outside the vortices.

**Fig. S4**

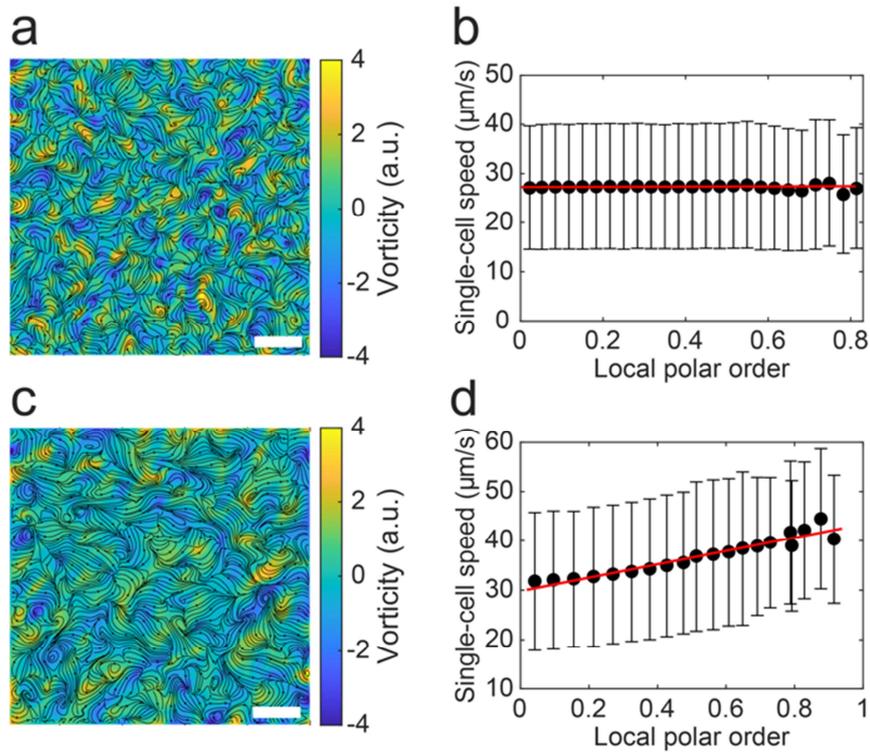

**Fig. S4. Behavior of quasi-2D bacterial active fluids below the critical cell density for developing ordered vortex lattice.** (**a**, **c**) Spatial distribution of instantaneous vorticity computed based on collective velocity field of bacterial active fluids (panel a: $2.0 \times 10^{10}$ cells/mL; panel c, $3.7 \times 10^{10}$ cells/mL). The vorticity fields are plotted in the same manner as in main text Fig. 1c. Both panels show disordered spatial distribution of vortices. Panels a and c share the same scale bar (500 μm). (**b**, **d**) Apparent single-cell speed plotted against local polar order (as a proxy of local collective active stress) in bacterial active fluids (panel b: $2.0 \times 10^{10}$ cells/mL; panel d, $3.7 \times 10^{10}$ cells/mL). In panel b, single-cell speed is almost independent of local polar order (i.e., the mobility enhancement coefficient $\beta \sim 0$). In panel d, single-cell speed is weakly correlated with local polar order ($\beta \approx 0.48$). Error bars in panel b and d indicate standard deviation (N>50 single-cell trajectories for each data point).

**Fig. S5**

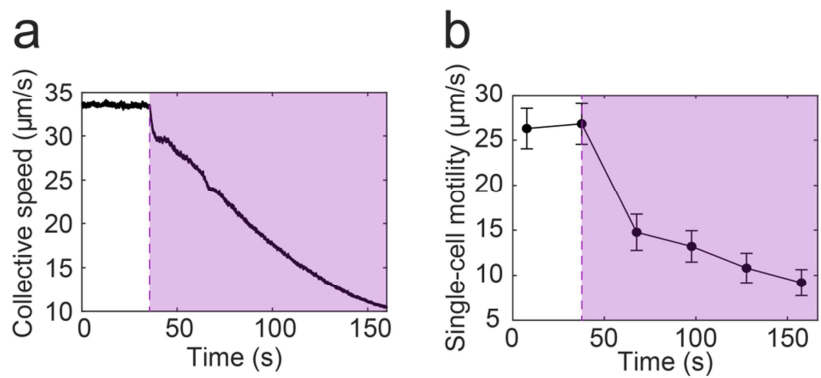

**Fig. S5. Suppression of bacterial motility by violet light illumination**. (**a**) Collective speed averaged over the entire field of view of the dense bacterial suspension was plotted as a function of time. The vortex lattice was illuminated by violet light at T = 35.5 s (violet dashed line) and the motility of all cells in the entire field of view was suppressed by violet light illumination. This figure is associated with Video S3. (**b**) Tuning single-cell motility of *S. marcescens* via violet-light illumination. Single-cell motility refers to the intrinsic speed cells in an isolated environment; it is different from the apparent single-cell speed measured in dense suspensions where cell's motion is affected by the mechanical environment. To measure single-cell motility, cells were extracted from dense *S. marcescens* suspensions that displayed the large-scale ordered vortex lattice, diluted to an appropriate density, and deposited on freshly made 0.6% LB agar surface to form a quasi-2D dilute bacterial suspension drop. *S. marcescens* cells in the prepared suspension drop were tracked in fluorescence microscopy through a 20x objective lens, and starting from T = 35.5 s the cells were continuously illuminated by 406 nm violet light (Methods). The environmental temperature was maintained at 30 °C with a custom-built temperature-control system (Methods). The speed of an isolated cell (i.e., the motility of the cell) at a specific time T was computed based on its trajectory tracked from (T-0.5) s to (T + 0.5) s; the speeds of isolated cells computed from (T-12.5) s to (T + 12.5) s were then averaged and taken to be the mean of single-cell motility at time T. Before violet-light illumination, the temporal variation of the speed of isolated cells is ~17%, suggesting that cells prepared from the same overnight culture should have a similar self-propulsion force. Data presented in this figure show that the average collective speed measured in the dense suspension (panel a) is proportional to single-cell motility (panel c) measured in isolated environment during violet light illumination. Therefore, the average collective speed is an appropriate proxy of single-cell motility.

**Fig. S6**

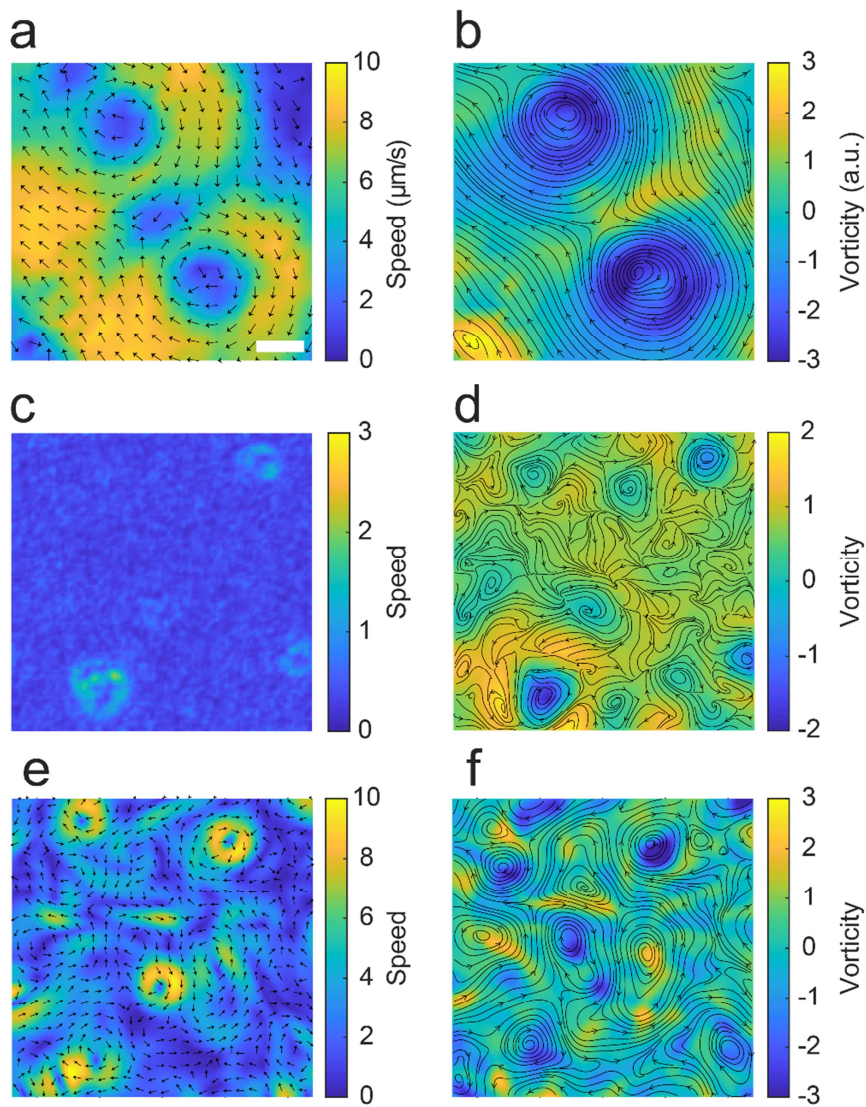

**Fig. S6. Sporadic vortices in quasi-2D dense *E. coli* suspensions and in numerical simulations**. (**a, b**) Time-averaged collective velocity field (panel a) and vorticity field (panel b) experimentally measured in a dense *E. coli* suspension (Methods) showing sporadic and disordered vortices. Data presented in the two panels were averaged over a duration of 10 s. In panel a, arrows and colormap represent collective velocity directions and magnitude, respectively, with the colorbar provided to the right, in unit of µm/s. The vorticity field shown in panel b was plotted in the same manner as in main text Fig. 1c. Panels a,b share the same scale bar, 500 µm. Also see Video S4 for the experimental video. (**c, d**) Time-averaged particle speed spatial distribution (panel c) and vorticity field (panel d) showing sporadic and disordered vortices in particle-based simulation with a low particle activity (simulation parameters: particle activity $f_0 = 7$ and mobility enhancement coefficient $\beta = 1.2$). Data presented in the two panels were averaged over a duration of 10 time units. Colormap in panel c represents particle speed, with the colorbar provided to the right. The vorticity field shown in panel d is plotted in the same manner as in main text Fig. 1c. (**e,f**) Time-averaged collective velocity field (panel e) and vorticity field (panel f) showing sporadic and disordered vortices in continuum modeling with a low activity (simulation parameters: activity $|S| = 2.6$ and mobility enhancement coefficient $\beta = 2.0$). Data presented in the two panels were averaged over a duration of 10 time units and plotted in the same manner as panels a,b. The simulation parameter sets in panels c,d and e,f were chosen from the boundary between the active turbulence state and the vortex lattice state in the phase maps shown in main text Fig. 4c and Fig. 4f, respectively.

**Fig. S7**

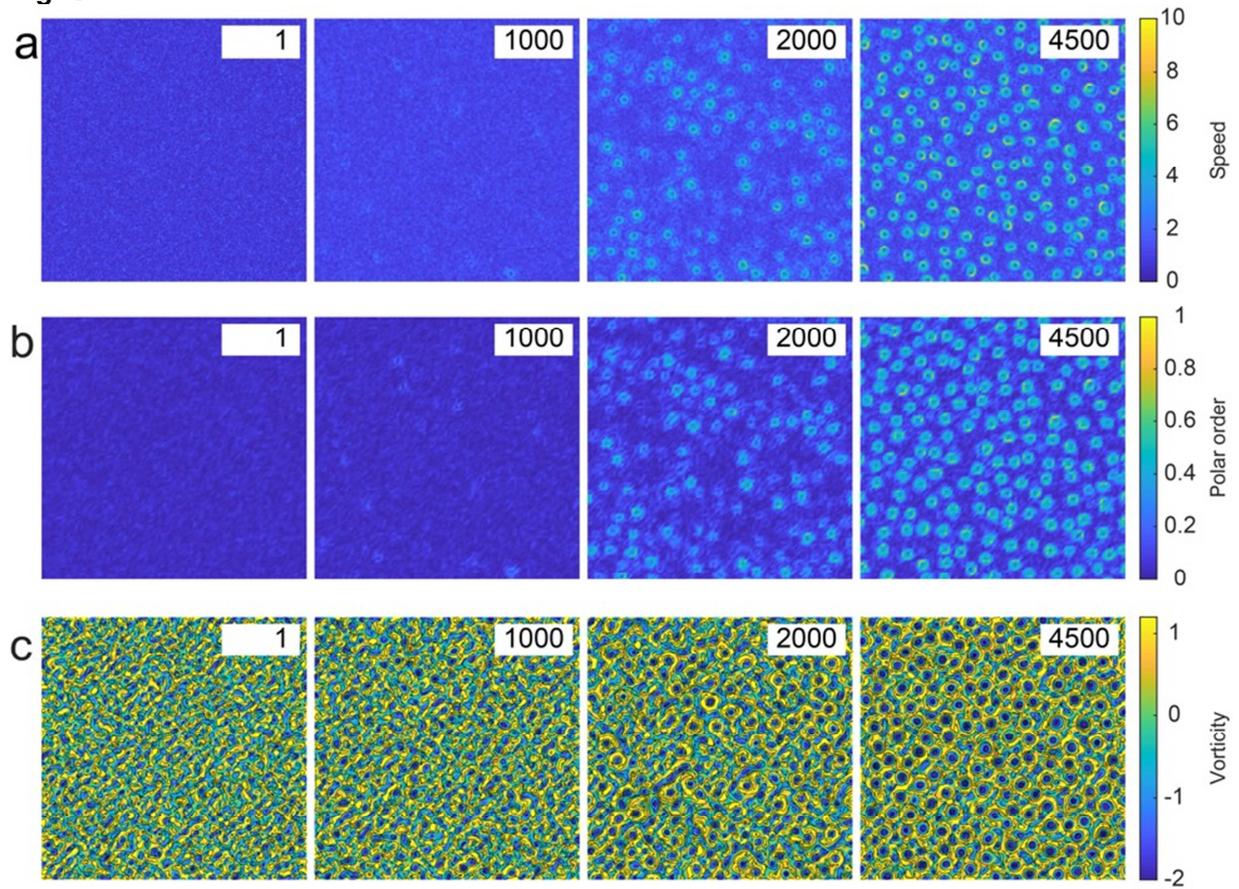

**Fig. S7. Emergence of vortex lattice pattern in the particle-based simulation**. Time sequence of spatial distributions of instantaneous collective speed (panel a), local polar order (panel b), and vorticity (panel c) during the emergence of ordered vortex lattice in the particle-based simulation. The magnitude of particle speed, local polar order and vorticity is indicated by the colorbars provided to the right of each panel. The vorticity field shown in panel c was plotted in the same manner as in main text Fig. 1c. The time stamp in each sub-panel indicates the elapsed time units in the simulation. Simulation parameters: particle activity $f_0 = 10$ and mobility enhancement coefficient $\beta = 1.2$. This figure is associated with main text Fig. 4d,e.

**Fig. S8**

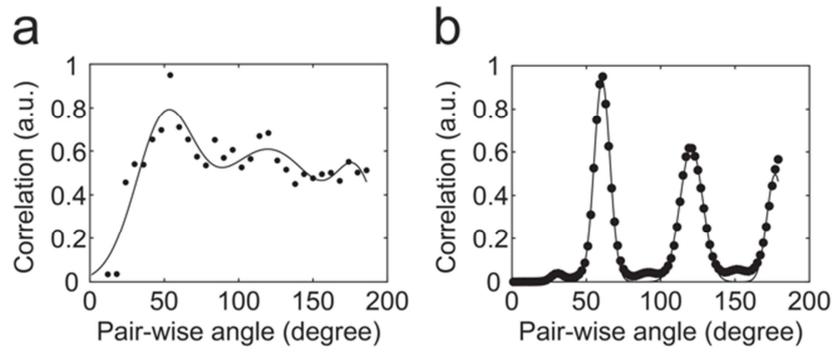

**Fig. S8. Triplet-distribution function in particle-based (panel a) and continuum (panel b) simulations.** Solid lines were obtained by fitting the data into a sum of three Gaussian functions (Methods). Simulation parameters were identical to those used in main text Fig. 4d,e and Fig. 4g,h, respectively: particle activity $f_0 = 10$ and mobility enhancement coefficient $\beta = 1.2$ in panel a; activity $|S| = 3.5$ and mobility enhancement coefficient $\beta = 2.0$ in panel b.

**Figure S9**

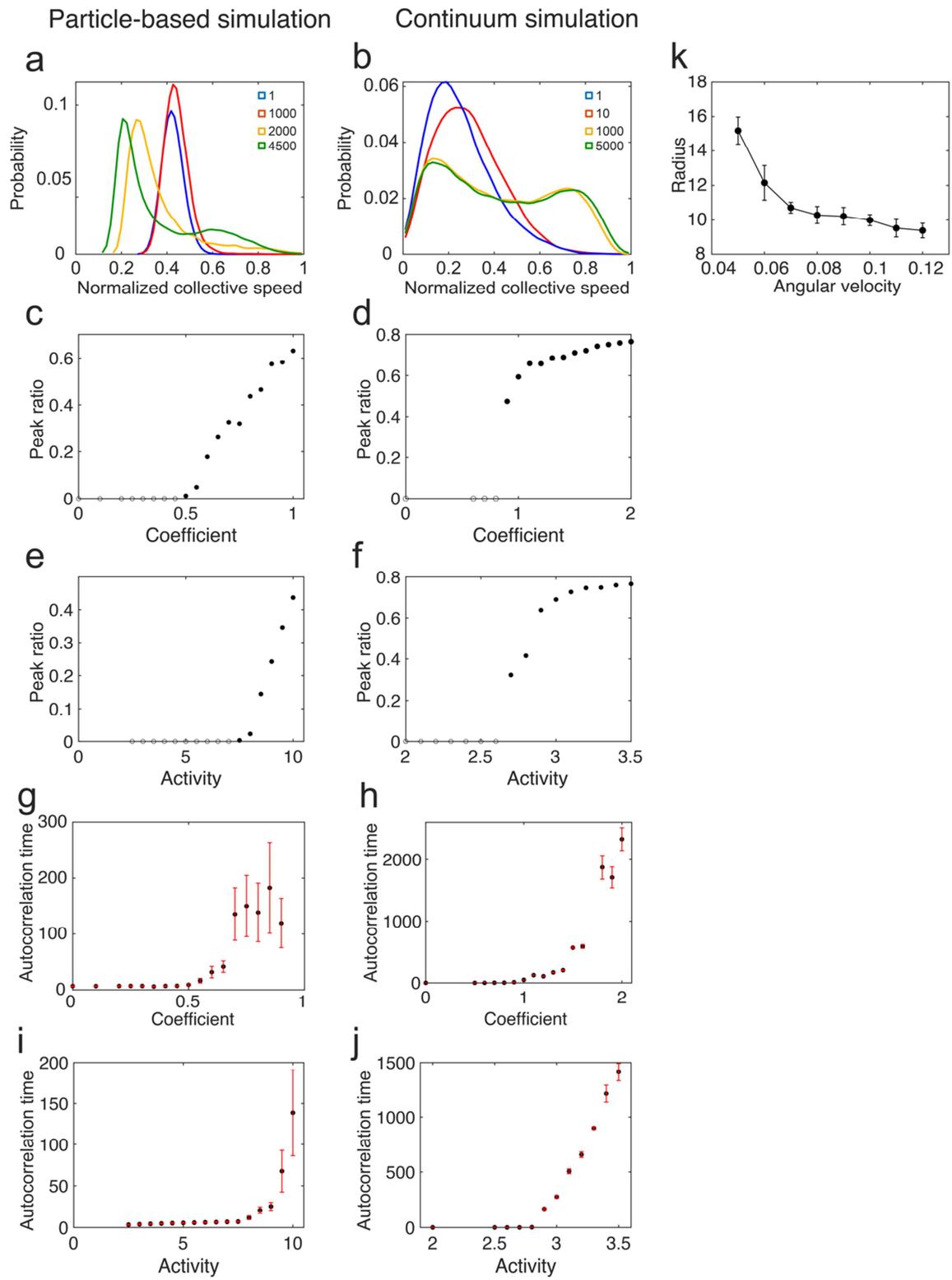

**Fig. S9. Characterization of signatures for the transition to stable vortex lattice state in particle-based and continuum modeling.** Bimodality of collective speed distribution and vortex lifetime are used as the signatures for the transition to stable vortex lattice state (see Fig. 3). Panels a, c, e, g, i: results from particle-based simulations; panels b, d, f, h, j: results from continuum simulations. (**a,b**) Probability distributions of normalized particle speed (panel a) and of collective speed (panel b) during the emergence of vortex lattice pattern in a representative simulation. The two panels should be compared to the experimental result in main text Fig. 3c. The colors represent distributions at different time units elapsed in the simulations: panel a, blue: 1, red: 1000, yellow: 2000 and green: 4500; panel b, blue: 1, red: 10, yellow: 1000 and green: 5000. Both panels show that the speed distribution transited from being unimodal to bimodal. Simulation parameters were identical to those used in main text Fig. 4d,e and Fig. 4g,h: particle activity $f_0 = 10$ and mobility enhancement coefficient $\beta = 1.2$ in panel a; activity $|S| = 3.5$ and mobility enhancement coefficient $\beta = 2.0$ in panel b. (**c,d**) Peak ratio in collective speed distribution (Methods) plotted against mobility enhancement coefficient in particle-based simulations (panel c) and in continuum simulations (panel d). The two panels should be compared to the experimental result in main text Fig. 3e, as mobility enhancement coefficient is cell density is positively correlated with cell density (see main text Fig. 4b). (**e,f**) Peak ratio in collective speed distribution plotted against the activity $f_0$ in particle-based simulations (panel e) and against the activity parameter $|S|$ in continuum simulations (panel f). The two panels should be compared to the experimental result in main text Fig. 3f. (**g,h**) Steady-state autocorrelation time of collective velocity (as a measure of vortex lifetime; Methods) plotted against mobility enhancement coefficient in particle-based simulations (panel g) and in continuum simulations (panel h). The two panels should be compared to the experimental result in main text Fig. 3d. (**i,j**) Steady-state autocorrelation time of collective velocity plotted against activity $f_0$ in particle-based simulations (panel i) and activity parameter $|S|$ in continuum simulations (panel j). In panel c-f, the solid circles represent data from bimodal speed distributions, while the empty circles correspond to zero peak ratio and represent data from unimodal speed distributions (i.e., the higher-speed peak has a height of zero). Simulation parameters: In particle-based simulations, activity $f_0 = 10$ and mobility enhancement coefficient $\beta$ from 0 to 1.0 for panels c,g; activity $f_0$ from 2 to 10 and $\beta = 0.8$ for panels e,i. In continuum simulations, activity $|S| = 3.5$ and mobility enhancement coefficient $\beta$ from 0 to 2 for panels d,h; $|S|$ from 2 to 3.5 and $\beta = 2.0$ for panels f,j. (**k**) Mean radius of vortices plotted against the mean angular velocity of particles in a simplified particle-based model, where the time evolution of particle angular velocity $\omega_i$ (Eq. [3] in Methods) is replaced by drawing its value from a distribution derived from the experimental curvature measurement in Fig. S11. The simplified model produces similar results in terms of vortex pattern emergence. By varying the mean angular velocity $\Omega = \langle \omega_i \rangle_i$ (while keeping the shape or standard deviation of the probability distribution for normalized $\omega_i$ the same), the simulation here shows that the mean radius of the vortices is negatively correlated with $\Omega$ (which sets the intrinsic curvature of particle motion). The vortex size is measured by thresholding the vorticity field derived from the collective velocity field (Methods). Error bars represent the standard deviation, N=20. Simulation parameters: $f_0 = 10$ and $\beta = 1.2$.

**Fig. S10**

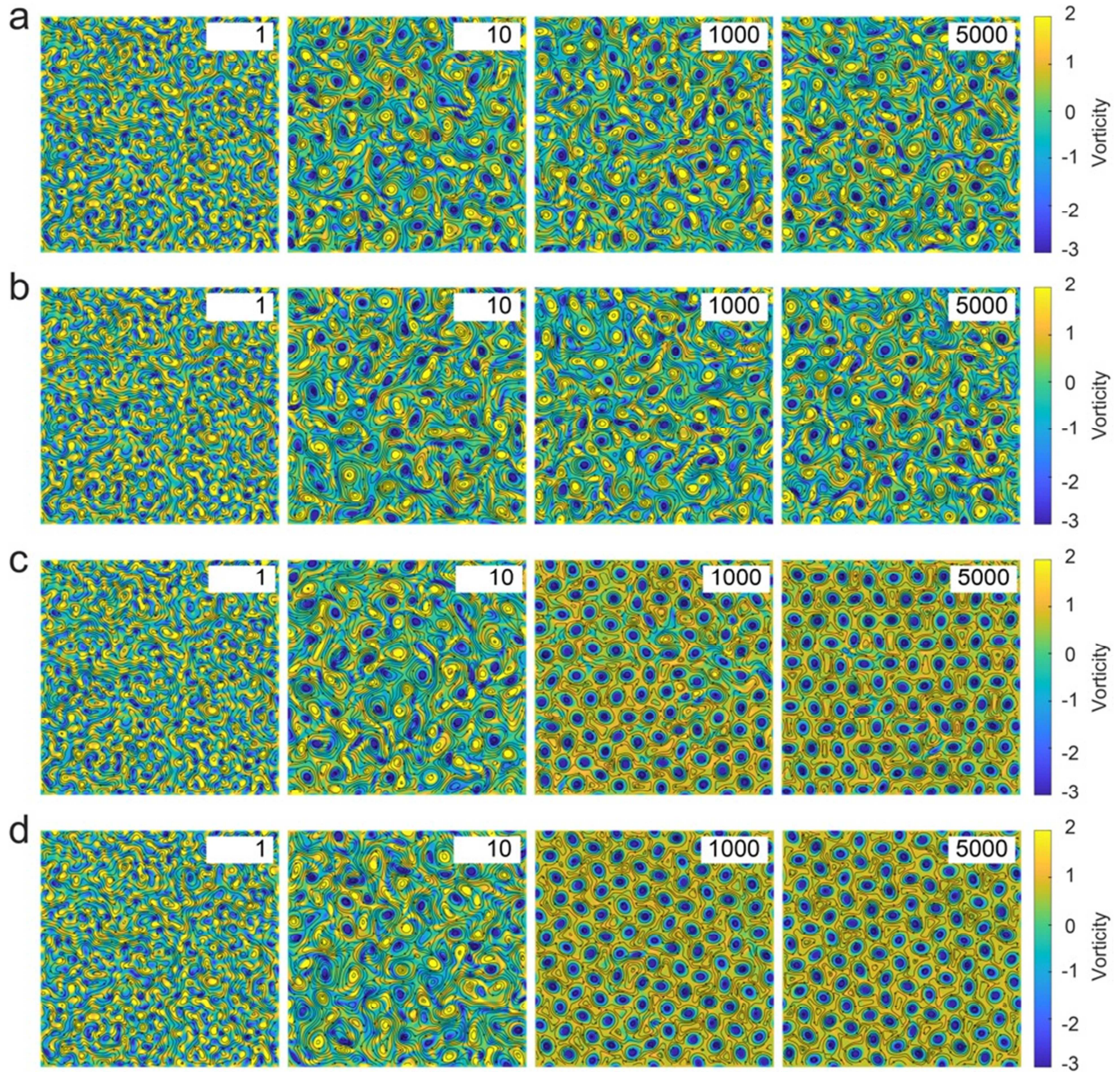

**Fig. S10. Emergence of collective motion patterns in the continuum model.** This figure shows the time sequences of vorticity field in numerical simulations of the continuum model with different mobility enhancement coefficient ($\beta = 0, 0.5, 1.0$ and $2.0$ for panels a, b, c and d, respectively; activity $|S| = 3.5$ for all panels). The time stamp in each sub-panel indicates the elapsed time units in the simulation. The vorticity fields are plotted in the same manner as in main text Fig. 1c.

**Fig. S11**

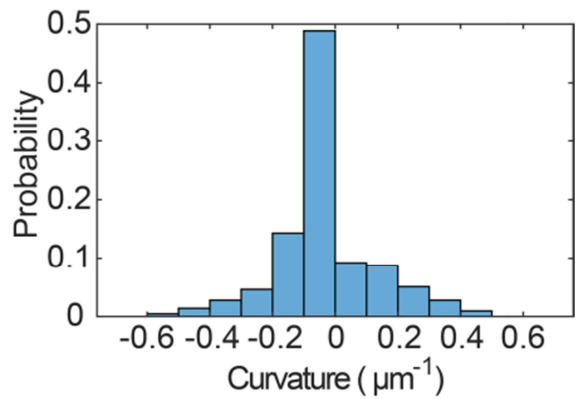

**Fig. S11. Motion bias of isolated *S. marcescens* cells swimming near a solid substrate.** This figure shows the probability distribution of the signed curvature of single-cell trajectories near a solid substrate, with the mean curvature being -0.023 µm$^{-1}$. The signed curvature of cell trajectories was computed based on 1-s segments of cell trajectories (positive: CCW; negative: CW). To obtain the distribution, cells were extracted from dense *S. marcescens* suspensions that displayed the large-scale ordered vortex lattice, diluted to an appropriate density, and deposited on freshly made 0.6% LB agar surface to form a quasi-2D dilute bacterial suspension drop. *S. marcescens* cells in the prepared suspension drop were tracked in fluorescence microscopy, while the environmental temperature was maintained at 30 °C with a custom-built temperature-control system (Methods).